\documentclass[12pt]{article}

\usepackage{preprint}
\usepackage{amssymb}
\usepackage{wasysym}
\usepackage{graphicx}
\usepackage{changebar}

\begin{document}

\hyphenation{cal-o-rim-e-ter}

\title{\bf Radiation hardness of the PIBETA detector components}

\author{E.~Frle\v{z}, T.~A.~Campbell, I.~J.~Carey, and D.~Po\v cani\'c \\[2ex]
 Department of Physics, University of Virginia,  \\[1ex]
 Charlottesville, VA~22904-4714, USA 
               }
\date{19 November 2002}
\maketitle
\thispagestyle{empty}\pagestyle{plain}

\begin{abstract}

We have examined long term changes in signal amplitude gain, energy
resolution and detection efficiency for the active components of the
PIBETA detector system.  Beam defining plastic scintillation counters
were operated in a $\sim$\,1\,MHz stopped $\pi^+$ beam for a period of
297 days, accumulating radiation doses of up to $2\cdot 10^6\,$rad.
Detectors in the charged particle tracking system---a pair of
cylindrical multi-wire proportional chambers and a thin plastic
scintillation barrel-shaped hodoscope array---were irradiated during
the same running period with an average dose of $\sim 4 \cdot
10^4\,$rad.  Individual CsI(undoped crystal) calorimeter detectors
received an average dose of $\sim$\,120\,rad, mainly from photons,
positrons and protons originating from $\pi^+$ hadronic interactions
as well as from $\pi^+$ and $\mu^+$ weak decays at rest in the active
target.

\vspace*{3ex}


\end{abstract}

\setlength\baselineskip{18pt} 

\section{Introduction}

The PIBETA collaboration has proposed a program of precise
measurements of rare $\pi$ and $\mu$ decays at the Paul Scherrer
Institute (PSI)~\cite{Poc91}, with particular emphasis on the pion
beta decay branching ratio, $\Gamma(\pi^+\rightarrow\pi^0e^+\nu_e)$.

\begin{figure}[hbt]
\noindent\hbox to \textwidth{\hfill
 \resizebox{0.95\textwidth}{!}
            {\includegraphics{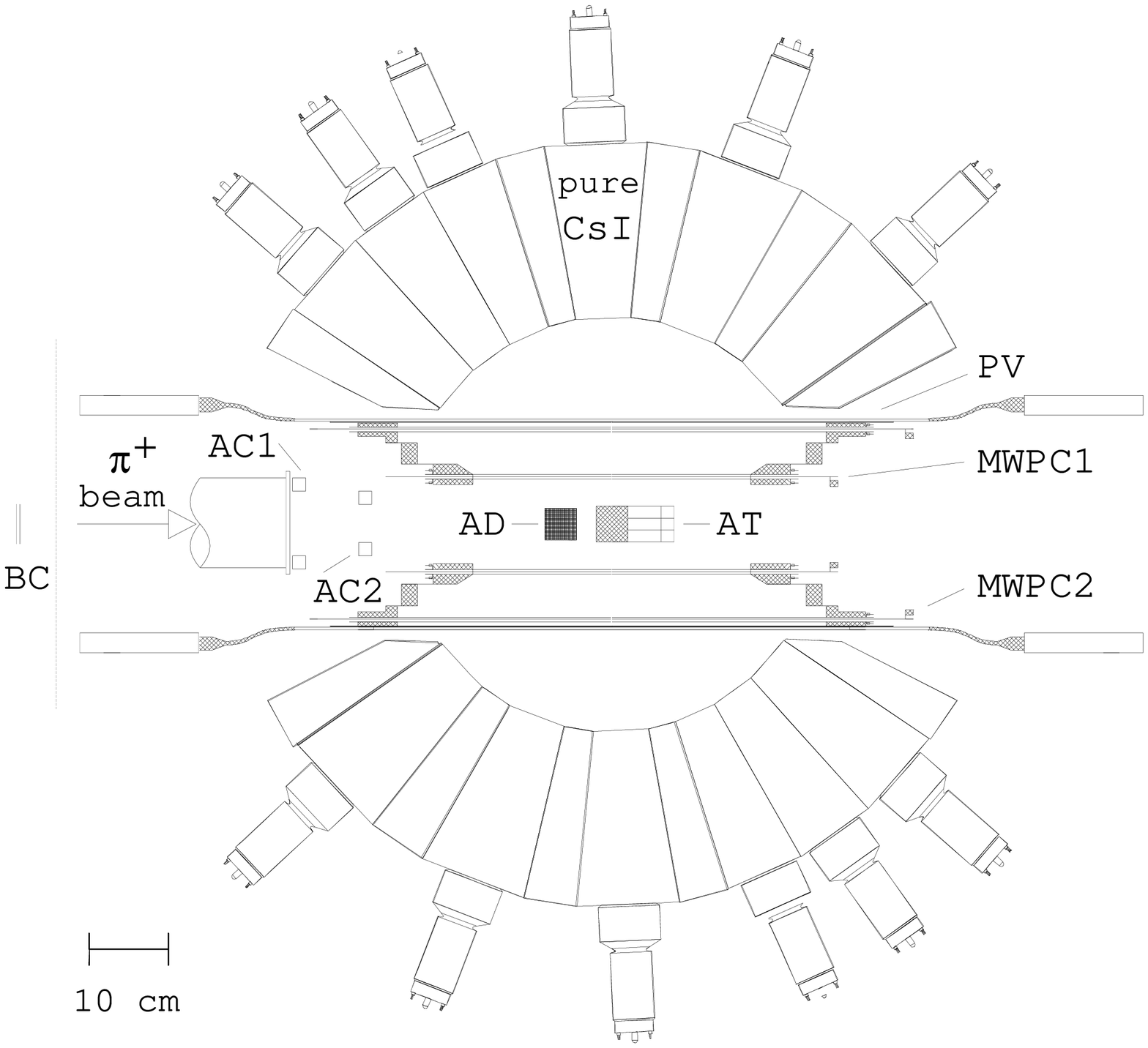}} \hfill}
\noindent
\begin{minipage}[b]{0.45\textwidth}
 \caption{(a) [above] Schematic cross section of the PIBETA apparatus
   showing the main components: beam entry counters (BC, AC1, AC2),
   active degrader (AD), active target (AT), wire chambers (MWPCs) and
   support, plastic veto (PV) detectors and PMTs, pure CsI calorimeter
   and PMTs.  (b) [right] Axial (beam) view of the central detector
   region showing the 9-element active target and the charged particle
   tracking detectors. \vspace*{3ex} }
\end{minipage}
\hfil
\resizebox{0.5\textwidth}{!}
             {\includegraphics{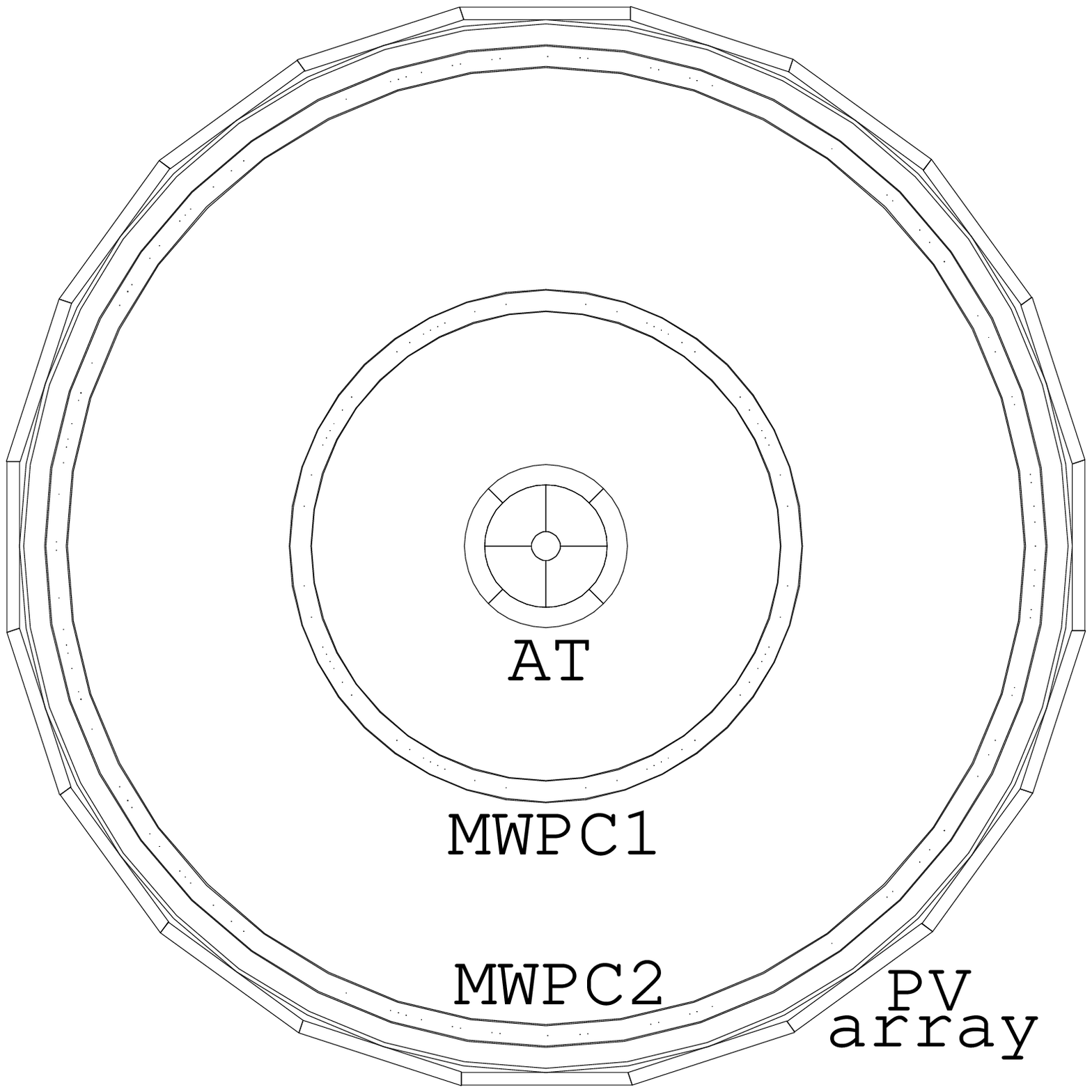}}
\label{fig1}
\end{figure}

The PIBETA apparatus is a large solid angle non-magnetic detector
optimized for detection of photons and electrons in the energy range
of 5--150$\,$MeV with high efficiency, energy resolution and solid
angle.  The main sensitive components of the apparatus, shown and
labeled in Fig.~\ref{fig1}, are:
\begin{itemize}
\item[(1)] BC, a thin forward beam counter placed approximately 4$\,$m 
upstream of the detector center, AC$_1$ and AC$_2$, two cylindrical
active collimators, AD, an active degrader, all made of plastic
scintillator and used for beam definition;
\item[(2)] AT, a 9-element segmented active plastic scintillator
target, used to stop the beam particles while simultaneously sampling
the lateral beam profile;
\item[(3)] MWPC$_1$ and MWPC$_2$, two concentric low-mass cylindrical
multi-wire proportional chambers for charged particle tracking,
surrounding the active target; 
\item[(4)] PV, a fast 20-bar segmented thin plastic scintillator
hodoscope, surrounding the MWPCs, used for particle identification;
\item[(5)] a 240-element fast high-resolution segmented spherical
pure-CsI shower calorimeter surrounding the target region and tracking
detectors, subtending a solid angle of $\sim 80\,$\% of $4\pi$;
\item[(6)] CV, a set of cosmic muon plastic scintillator veto counters
around the entire apparatus, not shown in Fig.~\ref{fig1}.
\end{itemize}

\bigskip

The detector components listed above, together with delay cables for
photomultiplier tube (PMT) analog signals, high voltage (HV) supplies
and cables, MWPC instrumentation and gas system, fast trigger
electronics, two front end computers (one for data acquisition, the
other for slow control), as well as the temperature control system,
are all mounted on a single platform that is moved as a single unit in
and out of the experimental area.  Thus, the detector can become fully
operational in less than 24 hours after the support platform is
precisely positioned with respect to the beam line, and electrical
power and Ethernet connections are made.

The building and testing of the detector components were completed in
1998, followed by the assembly and commissioning of the full detector
apparatus.  Data acquisition with the PIBETA detector started in the
second half of 1999, initially at a reduced pion stopping rate, as
planned.  Since then, the pion stopping rate was gradually increased
and the experiment ran during most of the available beam period in the
years 2000 and 2001 at $\sim 1\,$MHz $\pi^+$ stopping rate.

In all important respects the detector has met its design
specifications.  In this paper we report on the radiation resistance
and the temporal stability of the gain, energy resolution, and
detection efficiency of the most affected active elements of the
PIBETA detector listed above.

\clearpage

Radiation stability of the plastic scintillator detectors used in high
energy and nuclear physics experiments is one of their main
characteristics, and as such has been discussed in a voluminous body
of research and review papers.  Here we note a review paper by
G.~Marini~et~al.~\cite{Mar85} and
Refs.~\cite{Bro90,Bic91,RAD93,RAD95,Sen95,And01}, as well as
references therein.  These papers address the issue of experimentally
determining and improving the radiation hardness of plastic
scintillators.

Radiation hardness of pure (undoped) CsI scintillators has been
reported in Refs.\ \cite{Kob92,Roo92,Woo92,Eig92,Kob93}.

\section{Experimental Analysis}

The PIBETA measurements are performed in the $\pi E1$ channel at
PSI~\cite{Psi94}.  For this experiment the beam line is operated in
the high intensity, low momentum resolution mode.  Correspondingly, a
114 MeV/c $\pi^+$ beam tune has been developed with momentum spread of
$\Delta p/p\le 1.2\,$\% and maximum nominal $\pi^+$ beam intensity of
$I_\pi \simeq 2\cdot 10^6\,\pi^+$/s.

The spatial spread of the $\pi^+$ beam is restricted by a $10\,$cm
thick lead collimator with a \diameter\,7\,mm pin-hole located
3985$\,$mm upstream of the detector center.  Beam particles are first
registered in the 2$\,$mm thick plastic scintillator (BC) placed
directly downstream of the collimator.  Pions are subsequently slowed
down in the 40$\,$mm long active plastic degrader (AD), and stopped in
the active plastic target (AT) positioned in the center of the PIBETA
detector, Fig.~\ref{fig1}.

We have analyzed a total of 6213 production runs, for which data were
accumulated between 9 October 1999 and 11 December 2000.  This data
set comprises a total number of $1.4\cdot 10^{13}$ beam $\pi^+$'s
stopped in the active target.  The $e^+$ and $\mu^+$ beam
contaminations measured in the BC--AT time-of-flight spectrum are
small, $\sim 1.3\,$\% and $\sim 0.3\,$\%, respectively.  Therefore,
the in-beam detectors were exposed primarily to pions, while the AT
counters also received significant doses from the stopped pion decay
products: $\pi \to \mu \to e$.  Particle discrimination between the
positrons, photons and protons detected in the CsI calorimeter is
accomplished using the charged particle tracking detector components,
i.e., $\rm MWPC_{1,2}$ and PV, the plastic veto hodoscope.

All individual detector PMT analog signals are discriminated in
time-over-thresh\-old CAMAC modules and counted with CAMAC scaler
units read out every 10$\,$s.  The cumulative scaler counts are
updated at the end of every production run in the online database, as
well as saved in a computer disk file.

The most probable, as well as the average, energy deposited in each
detector element are calculated in a Monte Carlo (MC) simulation using
the standard detector description code GEANT3~\cite{Bru94}.  The GEANT
simulation also provided the average values of radiation exposure
throughout the detectors' volumes.

The total energy absorbed per unit detector mass exposed to radiation
comprises the received radiation dose.  The absorbed radiation dose
for each detector is commonly expressed in units of rad~\cite{Leo87},
corresponding to the energy absorption of 100\,erg/g.  The equivalent
SI unit is 1 Gray equaling 100\,rad ($\rm 1\,Gy=1\,$Joule/kg).  The
absorbed radiation doses for each PIBETA active detector element are
calculated using the experimental and MC data on ionizing particle
types, cumulative particle rates and exposed detector volumes.  The
PIBETA detector absorbed doses are listed in Tables~\ref{tab1} and
\ref{tab3}.

The absolute energy calibration as well as the shapes of experimental
deposited energy (ADC) spectra are well understood in the MC GEANT
simulations.  The peaks in the ADC spectra accumulated for each series
of 10 runs are fitted with Gaussian functions in offline analysis.
The means of the Gaussian functions determine the relative detector
gain factors, while the standard deviations of the lower parts of the
ADC spectra loosely reflect the detector energy resolutions.

The PMT bias high voltages were in general kept constant, except for
the CsI calorimeter PMTs.  The demand HV values were set in 1\,V steps
with an accuracy and reproducibility of $\simeq$\,1\,V, which
corresponds to an equivalent gain change of $<0.5\,$\%.  For the 220
non-peripheral CsI calorimeter detectors high voltages were adjusted
automatically by the online analysis computer program on a daily
basis.  Firm constraints were imposed on the location of the $\pi\to
e\nu$ energy peaks.  The latter were always forced to the normalized
67.8\,MeV value by changing the PMT high voltages and adjusting the
calorimeter software gains appropriately.

For a detector equipped with an $n$-stage PMT operating in the linear
domain, a normalized real gain change $g$, relating two different
settings 1 and 2 (at time $t_1$ and $t_2$, respectively), depends on
the ratio of the software gains $s_i$ and the corresponding high
voltages ${\rm HV}_i$:
\begin{equation}
   g = \frac{s_1}{s_2} \rm\cdot \left(\frac{HV_1}{HV_2}\right)^n\ .
                                                   \label{eq:gain}
\end{equation}
This equation relates the true gain of a scintillator detector at time
$t_2$ to its gain at time $t_1$.

\section{Plastic Scintillator Beam Counters}

\subsection{Forward Beam Counter (BC)}

The forward beam counter BC is the first detector placed right after
the beam-defining lead collimator.  This counter tags the beam
particles that have passed through the collimator.

The central part of the beam counter is a quadratic piece of BICRON
BC-400 plastic scintillator~\cite{Bic89} with dimensions
$25\times25\times2\,$mm$^3$.  The scintillator is optically coupled on
all four sides to four tapered acrylic lightguides.  One lightguide is
glued to the scintillator edge surface with BICRON BC-600 optical
cement.  The other three lightguides have an air gap coupling to the
scintillator.

Both the scintillator and the lightguides are mounted inside a
light-tight enclosure.  This detector-carrying box consists of an
aluminum frame with outer dimensions $150\times150\times50\,$mm$^3$,
covered by two thin aluminum windows, each $30\,\mu$m thick.  The Al
frame is attached to the lead collimator and keeps the counter
position fixed.  Feedthroughs at the four lateral sides of the box
hold the light guides and the magnetic shield cylinders of the
PMTs.  The box is therefore also used as a mount for the
photomultipliers, while keeping the scintillator counter light-tight
and protecting it mechanically.

Each of the four lightguides is air-gap coupled to a Hamamatsu R7400U
mini-PM tube. The stabilized active PMT voltage divider was specially
designed at the University of Virginia for high rate operation,
because typical counter rates are $>2\,$MHz.  The four analog PMT
signals are electronically summed in a NIM LCR 428F linear
Fan-in/fan-out unit set up in the experimental beam area.

The analog pulses from the beam counter are divided by a custom-made
passive splitter into two signals of equal amplitude.  One signal is
discriminated in a CAMAC discriminator module; discriminator output is
digitized in a FASTBUS Time-to-Digital Converter (TDC) as well as
counted with a CAMAC scaler unit.  The second signal is connected to a
FASTBUS Analog-to-Digital Converter (ADC), gated with a 100$\,$ns
event trigger gate.  Similar electronic logic is used with other
individual PIBETA detectors discussed in the following sections.

In order to suppress the background caused by detector hits not
associated with a $\pi$-stop event we use two active beam collimators
AC$_1$ and AC$_2$.  Both collimators are ring-shaped and are made of a
25.4$\,$mm thick BICRON BC-400 plastic scintillator.  AC$_1$, the
first (upstream) ring counter, has an outer diameter of 120$\,$mm and
an inner diameter of 50$\,$mm.  AC$_2$, closer to the detector center,
has the outer/inner diameter of 172/90$\,$mm, respectively.  These
dimensions were chosen such that the detectors subtend the various
beam tube elements and mechanical support parts of the detector
without intersecting the calculated envelope of the beam.

\begin{figure}[hbt]
\noindent\hbox to \textwidth{\hfill
    \resizebox{0.8\textwidth}{!}
                      {\includegraphics{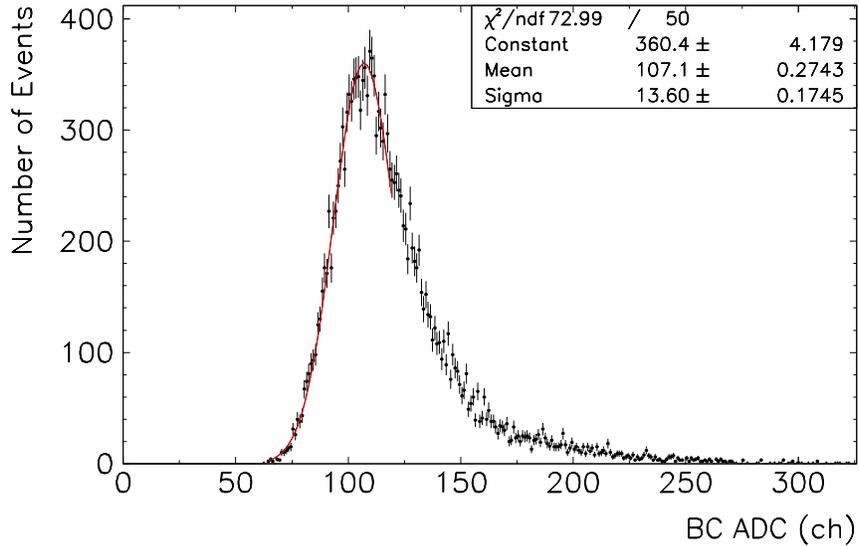}} \hfill}
\caption{Energy deposition (ADC signal charge) spectrum of 114\,MeV/c
$\pi^+$'s in BC, the thin forward beam counter.  Gaussian fit
parameters for the lower energy part of the peak are showed in the
statistics window.}
\label{fig2}
\end{figure}

The deposited energy (ADC signal charge) spectrum of the forward beam
counter recorded with the $\pi$-in-beam trigger is shown in
Fig.~\ref{fig2}.  Most probable energy deposition by a $\pi^+$ beam
particle corresponds to 0.70\,MeV, while the average absorbed energy
is 0.80\,MeV.  These values are calculated with a GEANT Monte Carlo
code using a 114\,MeV/c $\pi^+$ beam as an input.  The slope of the
low-energy ridge of the deposited energy spectrum in BC is determined
by a convolution of the Vavilov probability distribution and the PMT
photoelectron statistics.  Since the latter deteriorates with
radiation damage, the slope of the low-energy edge of the ADC spectrum
can be used as a rough indicator of the effects of radiation damage on
detector resolution, in the absence of a sharp monoenergetic line.  We
have used the same approach for the other PIBETA plastic scintillator
detectors discussed below: AD, AT, PV.

\begin{figure}[hbt]
\noindent\hbox to \textwidth{\hfill
   \resizebox{0.8\textwidth}{!}
                    {\includegraphics{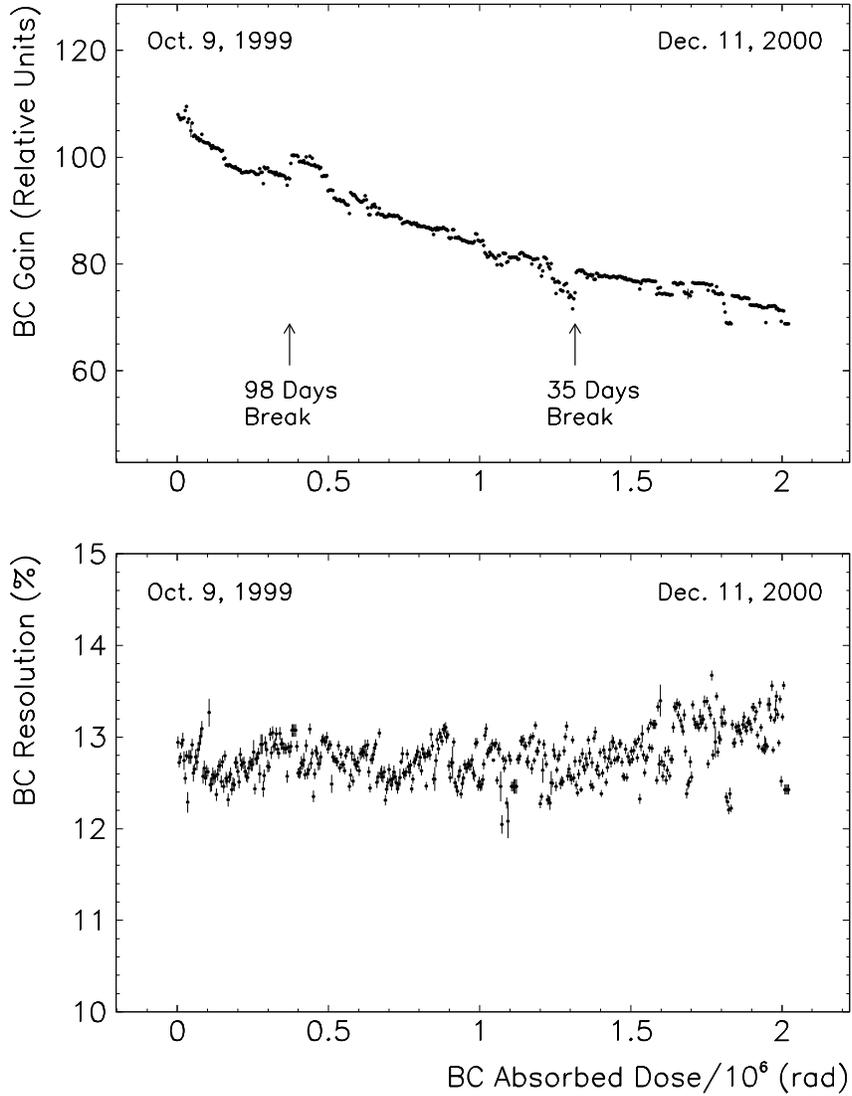}} \hfill}
\vspace*{-4ex}
\caption{Top panel: the change in gain of BC, the forward beam
counter, as a function of the absorbed radiation dose in Mrad.  Bottom
panel: fractional energy resolution of the BC detector as a function
of the cumulative radiation dose in Mrad.}
\label{fig3}
\end{figure}

Fig.~\ref{fig3} depicts the changes in the BC detector gain and
fractional energy resolution as the detector became damaged by
radiation over time.  It is evident from Fig.~\ref{fig3} that the gain
of the BC counter decreased approximately linearly over the period of
one calendar year due to the radiation exposure of 2 Mrad.  The
decrease in gain was $\simeq 20\,$\%/Mrad.  The fractional energy
resolution for through-going pions was also slightly degraded,
changing from 13.1\,\% to 13.9\,\%, as shown in the bottom panel of
Fig.~\ref{fig3}.  Two beam breaks in the radiation exposure, totaling
98 days and 35 days, respectively, resulted in a partial beam counter
recuperation and gain increases of $\simeq$\,10--15\,\%.  The
annealing effect on the counter resolution is smaller, though still
measurable.

\subsection{Active Degrader (AD)}

The active degrader counter is made of BICRON BC-400 plastic
scintillating material in the shape of a cup whose axis is aligned
with the beam.  The AD fits inside MWPC$_1$ which has an inner
clearance of \diameter\,90\,mm.  The AD outer surfaces are slanted and
connect to four acrylic lightguides.  This geometry brings the PMTs
out of the inner region, at the same time ensuring that the active
part of AD (cup ``bottom'') covers the whole target area (40\,mm
diameter).  The angle and wall thickness of the outer part of the AD
are chosen such that a particle traveling parallel to the beam axis
transverses the same thickness of 30\,mm.  The angle is kept small in
order to improve light collection efficiency.  Each of the four
fishtail lightguides ends in a rectangular cross section of $6 \times
6$\,mm$^2$, which matches the round window area of the Hamamatsu
R7400U PMT.  The tubes are mounted with an air gap coupling.  The four
AD PMT analog signals are summed in a NIM LCR 428F linear
Fan-in/fan-out unit.

\begin{figure}[hbt]
\noindent\hbox to \textwidth{\hfill
   \resizebox{0.8\textwidth}{!}
                      {\includegraphics{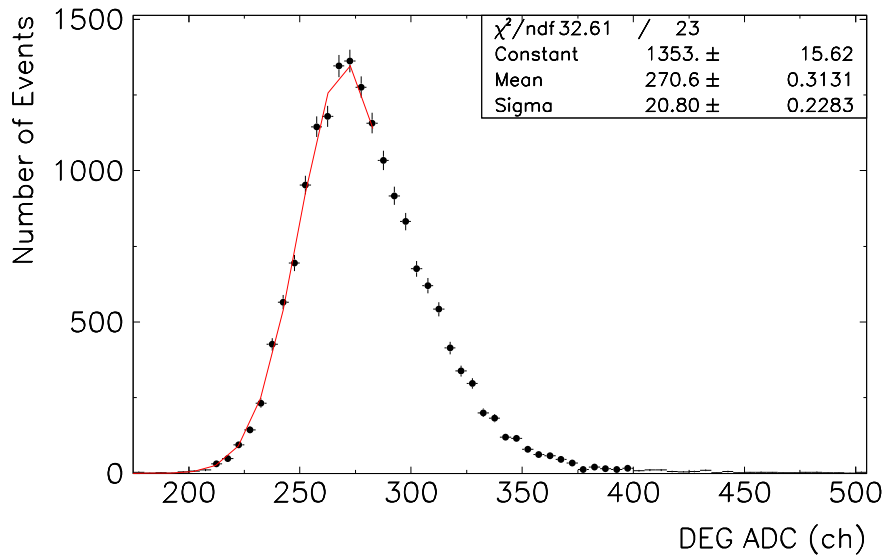}} \hfill}
\caption{ADC charge spectrum of the moderated $\pi^+$'s in the AD
counter.  A Gaussian fit (solid curve) is superimposed on the data
histogram (points).}
\label{fig4}
\end{figure}

\begin{figure}[hbt]
\noindent\hbox to \textwidth{\hfill
   \resizebox{0.8\textwidth}{!}
                      {\includegraphics{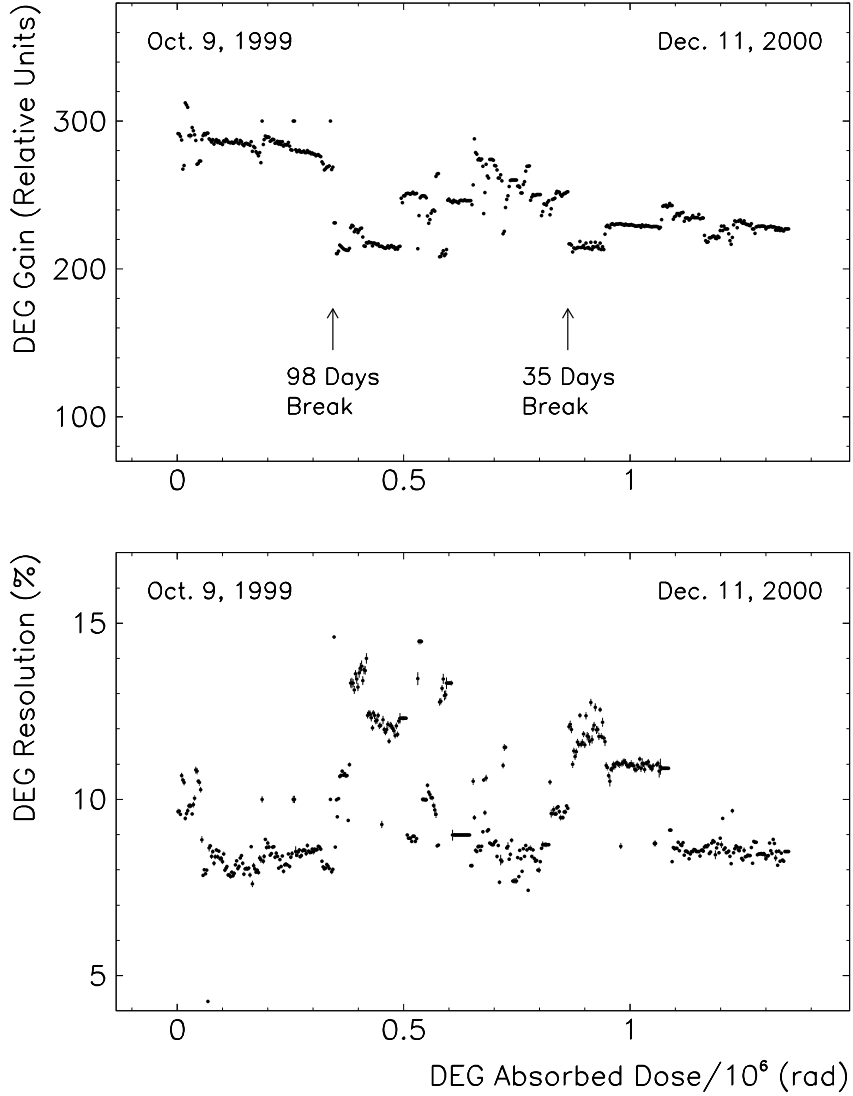}} \hfill}
\caption{Top panel: the gain of the AD counter as a function of the
absorbed radiation dose in Mrad.  Bottom panel: the energy resolution
of the active degrader as a function of the cumulative radiation dose
in Mrad.}
\label{fig5}
\end{figure}

The deposited energy (ADC charge) spectrum of the active degrader
counter taken with the $\pi$-in-beam trigger is shown in
Fig.~\ref{fig4}.  The lower energy part of the deposited energy
spectrum is again fitted well with a Gaussian function.  Most probable
energy deposition of the 114\,MeV/c incident $\pi^+$ beam is
calculated with the GEANT code to be equal to 13.0\,MeV, while the
average energy absorbed is 13.8\,MeV.  During the period under
analysis the degrader counter was irradiated with a dose of 1.4\,Mrad.
The changes in the relative counter gain and energy resolution are
shown in the two panels of Fig.~\ref{fig5}.  Both variables are
subject to short-term variations, mostly due to changes in the beam
tune.  There is, however, an average linear trend in the scintillator
gain factor corresponding to 15\,\%/Mrad decrease.  The temporal
degradation in the AD energy resolution is much smaller but still
noticeable.

\clearpage

\subsection{Active Stopping Target (AT)}

The PIBETA active target, AT, is a cylindrical plastic scintillator
counter 50$\,$mm long with a diameter of 40$\,$mm.  The counter is
segmented into 9 elements, as shown schematically in Fig.~\ref{fig1}b.
A \diameter\,7\,mm central cylinder is surrounded by 4 identical
tubular segments that compose an inner ring with outer
\diameter\,30.0$\,$mm.  The second, outer target ring is made of 4
tubular segments rotated by 45$^\circ$ with respect to the inner ring
(cf.\ Fig.~\ref{fig1}b).  The 9 pieces are wrapped individually in
aluminized Mylar foil isolating them optically from each other.  The
segments are pressed together and the whole modular assembly is
wrapped with black plastic tape.  Each target element is acting as an
independent counter, viewed by a miniature (8$\,$mm photo-cathode)
Hamamatsu R7400U photomultiplier tube via a 60$\,$mm long, tapered
acrylic lightguide.

\begin{figure}[hbt]
\noindent\hbox to \textwidth{\hfill
   \resizebox{0.8\textwidth}{!}
                  {\includegraphics{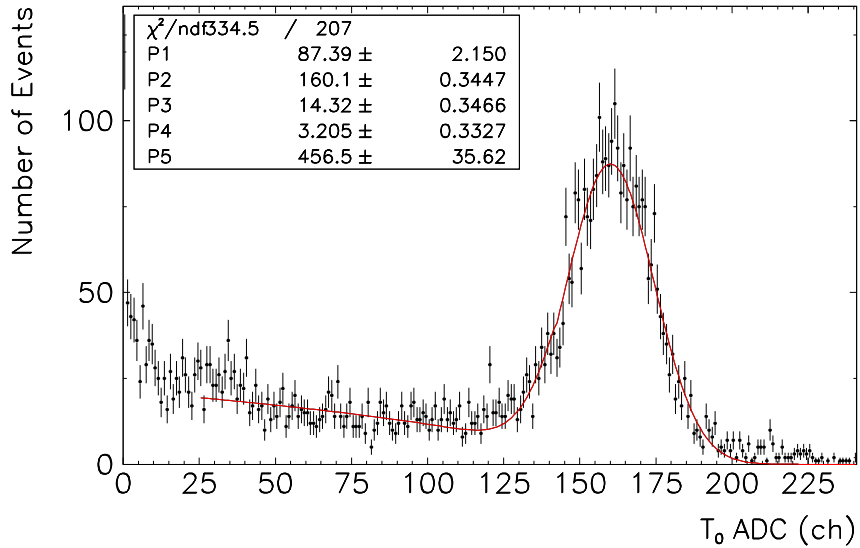}}  \hfill}
\caption{Deposited energy spectrum of stopping pions in the central
active target segment.  Parameters P1--P3 correspond to a Gaussian
function, the values P4 and P5 describe a square-root tail function
fit.}
\label{fig6}
\end{figure}

The uncalibrated deposited energy (ADC) spectrum of the central target
segment $\rm T_0$ recorded with the $\pi$-in-beam trigger is depicted
in Fig.~\ref{fig6}.  The events in which the stopping pion deposits
its full energy in the target element are clearly distinguished in the
high-energy peak.  The peak is well described by a Gaussian form with
a square-root function tail.  The GEANT simulation predicts a peak
position at 27.6\,MeV, and the average energy deposited in the central
target segment T$_0$ of 16.6\,MeV.  The events with a stopping
particle sharing energy between two or more target segments due to
scattering or to beam divergence, fall into the low energy tail.  The
same low energy tail is also populated by positrons from delayed weak
pion and muon decays.

\begin{figure}[hbt]
\noindent\hbox to \textwidth{\hfill
   \resizebox{0.8\textwidth}{!}
                {\includegraphics{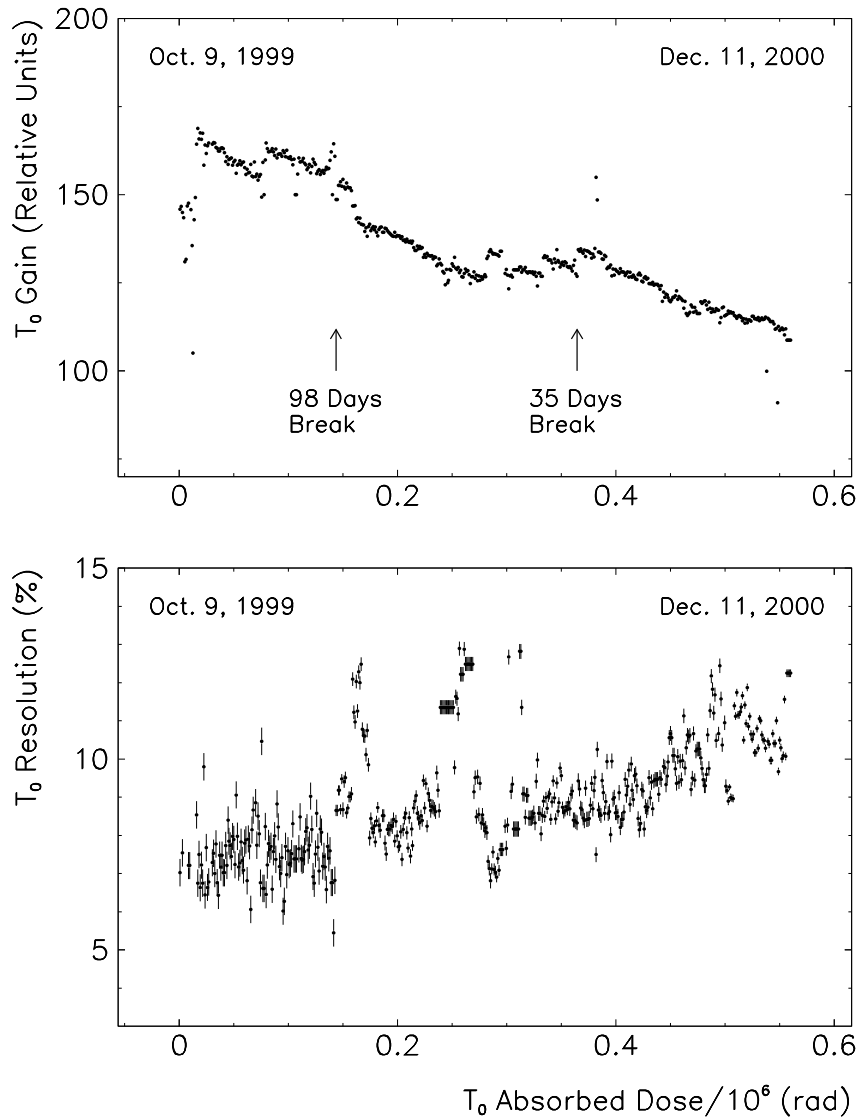}}  \hfill}
\caption{Top panel: the gain of AT$_0$, the central target segment, as
a function of the absorbed radiation dose in Mrad.  Bottom panel: the
energy resolution of the AT$_0$ counter as a function of the
cumulative radiation dose in Mrad.}
\label{fig7}
\end{figure}

In Table~\ref{tab1} we give the starting and final gains and
resolutions for the three representative target segments.  The
measured temporal variation in the gain factor and energy resolution
of the central target segment over the 15 month period is depicted in
Fig.~\ref{fig7}.  The cumulative absorbed radiation dose is calculated
to be 0.5\,Mrad.  A considerable fall-off in the gain of the T$_0$
scintillator is evident in the figure: $\simeq 52\,$\%/Mrad.
\clearpage
\noindent  As in
the AD counter, the energy resolution is subject to short-term
variations caused by changes in the beam spot geometry.  Over the
analyzed period, though, the resolution was degraded by a full 80\,\%.
This level of radiation damage makes it necessary to make available
identical copies of the AT detector for periodic replacement.  The
radiation damaged PIBETA target assembly has been replaced annually
with new target detectors.

\section{Multiwire Proportional Chambers (MWPCs)}

MWPC$_1$ and MWPC$_2$ are a pair of cylindrical multiwire proportional
chambers, each with one anode wire plane along the beam direction, and
two cathode strip planes in stereoscopic geometry.  A general
description of the design and operation of the ``Dubna''-type
cylindrical chambers is given in Ref.~\cite{Kra94}.  The PIBETA
detector cylindrical MWPCs are described in more detail in
Refs.~\cite{Kar98,Kar99}.  In this section we analyze the chamber
detection efficiencies measured with minimum ionizing particles.

The detection efficiency $\epsilon$ of MWPC$_1$ for minimum ionizing
particles (MIPs) can be measured using copious $\mu^+\to
e^+\nu\bar{\nu}$ (Michel) positrons emanating from the stopping
target:
\begin{equation}
   \epsilon_{\rm MWPC_1}
       = \frac{N(\rm AT \cdot MWPC_1\cdot MWPC_2\cdot PV \cdot CsI)}
           {N(\rm AT\cdot MWPC_2\cdot PV\cdot CsI)}\ ,
                                                  \label{eq:eff_mwpc}
\end{equation}
where the $N$'s represent the number of Michel events for which all
the detectors in the parenthesis register coincident hits above their
discriminator threshold.  The CsI calorimeter signal is discriminated
with a low threshold level ($\simeq 5\,$MeV), while a software cut on
the PV deposited energy spectrum selects MIP events (Fig.~\ref{fig8}).
An equivalent expression, obtained by substituting 2 for the index 1,
holds for MWPC$_2$.  
The average MWPC detection efficiencies at the
$\simeq 1\,$MHz $\pi^+$ beam stopping rate are $>95\,$\% and $>98\,$\%
for the inner and outer chambers, respectively.  Offline analysis of
the individual chamber efficiencies for the period under
consideration, which involved $\sim 1.4\cdot 10^{13}$ (mostly minimum
ionizing) particles, found no degradation within the measurement
uncertainty (Table~\ref{tab3}).

\section{Plastic Scintillator Hodoscope (PV)}

The PV (plastic veto) hodoscope is located inside the CsI calorimeter
and surrounds the two concentric wire chambers, as shown in
Fig.~\ref{fig1}.  The hodoscope array consists of 20 thin independent
plastic scintillator bars that are arranged to form a complete
cylinder with a $129\,$mm radius and a 598$\,$mm long axis that
coincides with the beam line and the target axis.  The PV hodoscope
covers the entire geometrical solid angle subtended by the CsI
calorimeter as seen from the target center.

A single PV detector element consists of four main components: (1)
thin plastic scintillator bar, (2) two lightguides, one at each end,
(3) two photomultiplier tubes, one per lightguide, and (4) aluminized
Mylar wrapping.

The dimensions of the individual plastic bars made of BC-400
scintillator are $3.18\times41.9\times598\,$mm$^3$.  Each thin
plastic scintillator bar, along with the attached lightguides, is
wrapped in 0.25$\,\mu$m thick aluminized Mylar foil to optically
separate it from the adjacent bars, and to provide a reflective
surface that increases the amount of light reaching the phototubes.
Scintillation light is viewed by two Burle Industries S83062E
photomultiplier tubes, one on each end. These tubes are
\diameter\,28\,mm head-on fast PMTs with 10 stages. 

The hodoscope array is supported from the inside by a $530\,$mm long
carbon fiber cylinder with a total thickness of $1\,$mm, corresponding
to $5.3\cdot 10^{-3}$ radiation lengths.  The hodoscope modules are
kept tight around the support cylinder by a helix-wound, thin plastic
string tensioned at the extreme ends of the detector stand.

\begin{figure}[hbt]
\noindent\hbox to \textwidth{\hfill
   \resizebox{0.8\textwidth}{!}
                {\includegraphics{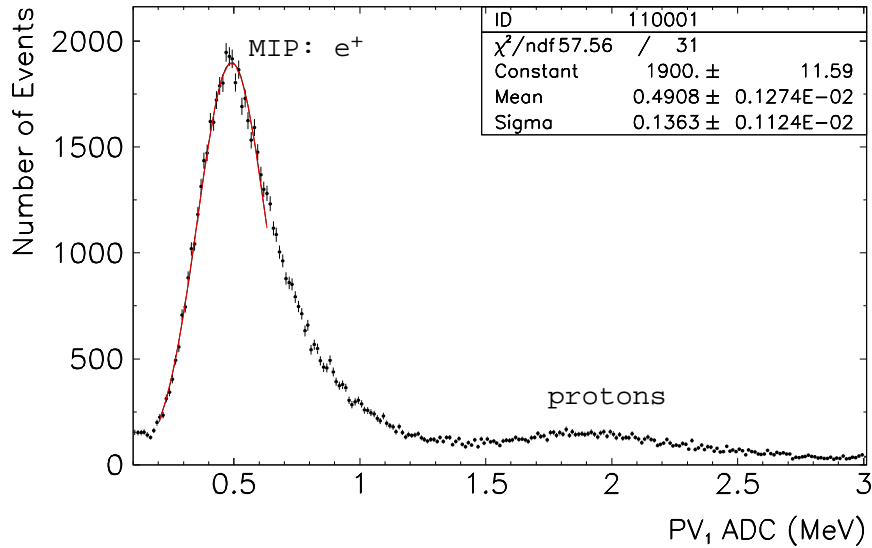}}  \hfill} \vspace*{-4ex}
\caption{Calibrated deposited energy (ADC charge) spectrum of
minimum-ionizing particles (mostly $e^+$'s, lower energy peak at
$\simeq$\,0.5\,MeV) and protons (peak at $\simeq$\,2.0\,MeV) in a
single representative detector element of the PV hodoscope.}
\label{fig8}
\end{figure}

The PV charged particle detection efficiency is evaluated separately
for mi\-ni\-mum-ion\-izing positrons with total energy above 5$\,$MeV
and for non-relativistic protons with kinetic energy in the range
10-150$\,$MeV.  Fig.~\ref{fig8} shows the measured energy spectrum of
positrons and protons in the PV detector, corrected for the angle of
incidence.

At the nominal $\pi^+$ stopping rate the individual plastic
scintillator phototubes are counting at the rate of $\sim 130\,$kHz,
while the entire PV hodoscope is taking $\sim 0.88\,$M hits/sec.  The
total radiation dose absorbed by the PV hodoscope array throughout the
analyzed period is 40\,krad.

A charged particle track is defined by the coincident hits in the
active target, AT, two wire chambers, MWPC$_1$ and MWPC$_2$, and the
CsI calorimeter, CsI.  The PV detection efficiency is defined as the
following ratio:
\begin{equation}
   \epsilon_{\rm PV}
    = \frac{N(\rm AT\cdot MWPC_1\cdot MWPC_2\cdot PV\cdot CsI)}
        {N(\rm AT\cdot MWPC_1\cdot MWPC_2\cdot CsI)}\ ,
                                                \label{eq:eff_pv}
\end{equation}
where each $N$ represents the number of events satisfying the
condition in the parenthesis.  The average PV detection efficiency
measured during the detector commissioning period was $\epsilon_{\rm
PV} \ge 99.0\,$\% (Table~\ref{tab3}).  No significant long term
change in the PV detection efficiency was observed.

\begin{figure}[hbt]
\noindent\hbox to \textwidth{\hfill
   \resizebox{0.8\textwidth}{!}
                {\includegraphics{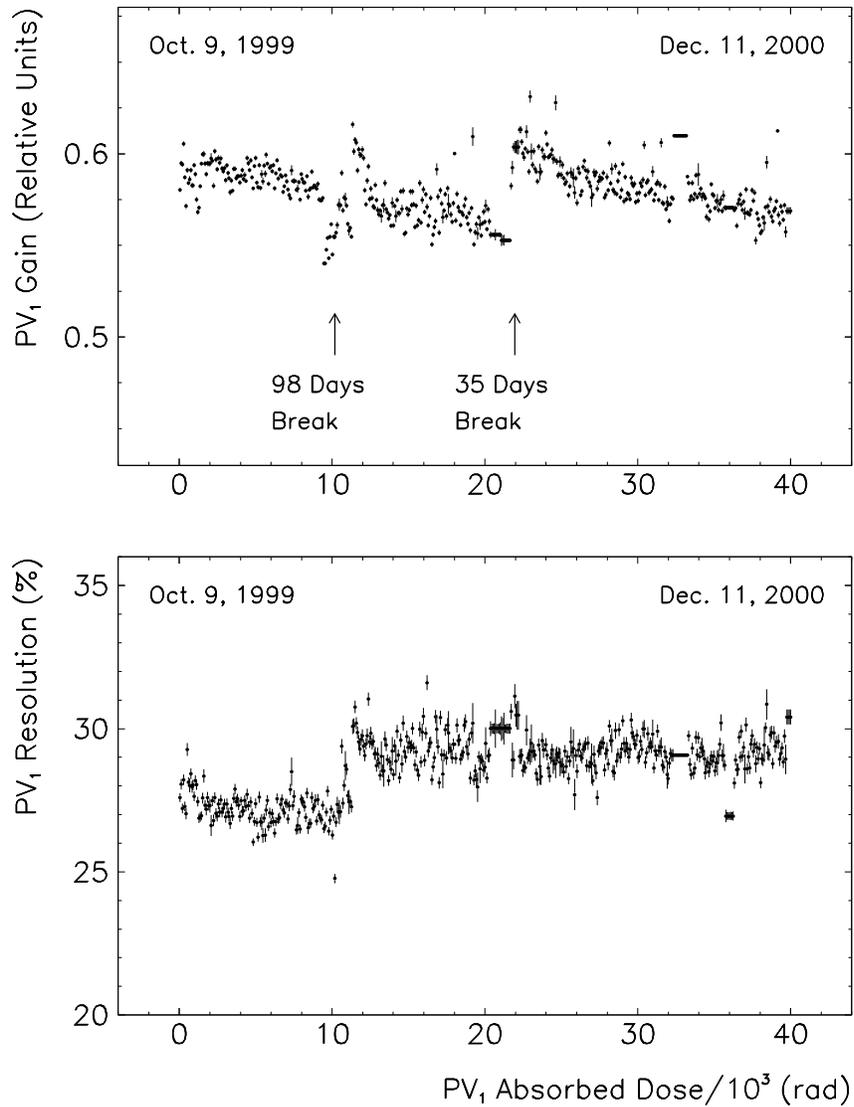}}  \hfill}
\caption{Top panel: the gain of the PV counter number~1 as a function
of the absorbed radiation dose in krad.  Bottom panel: fractional
energy resolution of the forward PV counter number~1 as a function of
the cumulative radiation dose in krad.  The discontinuity in the
fractional resolution plot around 10\,krad is related to a servicing
operation.}
\label{fig9}
\end{figure}

As for the previously discussed detectors, Table~\ref{tab1} lists the
starting and final gains and resolutions for two representative PV
detectors.  The temporal dependence of the detector light output for
PV bar number 1 is shown in Fig.~\ref{fig9}.  Its gain factor dropped
by 0.25\,\%/krad, giving a cumulative degradation over the production
period of 10\,\%.  The energy resolution on the other hand shows a
barely measurable change.

The charged particle tracking system, combining the information from
the PV hodoscope and the MWPC pair achieves MIP tracking inefficiency
in the range $(1.0\pm 0.2)\cdot 10^{-5}$\, while operating in the
environment of $\sim 1\,$MHz stopped $\pi^+$'s in the target.

\section{Pure CsI Calorimeter Modules}

The shower calorimeter lies at the heart of the PIBETA detector.  Pure
(undoped) CsI single crystal~(Refs.~\cite{Kob87,Kub88a,Kub88b}) was
chosen as the calorimeter material.

The PIBETA calorimeter consists of 240 pure CsI crystals.  Spherical
geometry is obtained by the ten-frequency class II geodesic
triangulation of an icosahedron~\cite{Ken76}.  The chosen geodesic
division results in 220 hexagonal and pentagonal truncated pyramids
covering a total solid angle of $0.77\times 4\pi\,$sr.  Additional 20
crystals surround two detector openings for the beam entry and
detector access and act as electromagnetic shower leakage vetoes.  The
inner radius of the crystal ball is 260$\,$mm, and the axial module
length is 220$\,$mm, corresponding to 12 radiation lengths, as
$X_0(\rm CsI) = 18.5\,$mm~\cite{PDG}.  

\clearpage

There are nine different
detector module shapes: four irregular hexagonal truncated pyramids
(labeled Hex-A, Hex-B, Hex-C, and Hex-D, respectively), one
regular pentagonal (Pent) and two irregular half-hexagonal truncated
pyramids (Hex-D1 and Hex-D2), plus two trapezohedrons which function
as calorimeter vetoes (Vet-1 and Vet-2).  The volumes of the PIBETA
CsI detector modules vary from 797$\,$cm$^3$ (Hex-D1/2) to
1718$\,$cm$^3$ (Hex-C).

The first 25 PIBETA crystals were manufactured in the Bicron
Corporation facility in Newbury, Ohio~\cite{Bic89}.  The remaining 215
CsI scintillators were grown in the Institute for Single Crystals in
Harkov (AMCRYS), Ukraine.  Preliminary quality control, including the
optical and mechanical crystal properties, was performed at the
production sites.

After completing physical measurements, all crystal surfaces were
hand-polished with a mixture of 0.2\,$\mu$m aluminum oxide powder and
etylenglycol.  The surfaces of each CsI crystal were then painted with
a special organo-silicon mixture.  The wavelength-shifting lacquer
developed by the Harkov Single Crystals Research Institute~\cite{xxx}
provides an optical treatment of crystal surfaces superior to more
common matting or wrapping treatments.  The lacquer contains a
wavelength-shifting ladder organosilicon copolymer with the chemical
composition PPO+POPOP+COUM.1, where PPO is
2.5--Di\-phe\-nyl\-ox\-a\-zole, POPOP represents
1.4-Di-2-(5-Phe\-nyl\-ox\-a\-zolile-Benzene) and COUM.1 is
7-Diethylamine-4-Me\-thyl\-cou\-ma\-rin~\cite{hand71}.

EMI 9822QKB 10-stage fast photomultipliers~\cite{EMI} with
\diameter\,75\,mm end windows are attached to the back faces of
hexagonal and pentagonal CsI crystals using a 300\,$\mu$m layer of
silicone Sylgard 184 elastomer (Dow Corning RTV silicone rubber plus
catalyst).  The resulting crystal--photomultiplier tube couplings are
strong and permanent, but can be broken by application of a
substantial tangential force.  Smaller half-hexagonal and trapezial
detector modules are equipped with \diameter\,46\,mm 10-stage EMI
9211QKA phototubes~\cite{EMI}.  Both photocathodes have quartz windows
transmitting photons down to $175\,$nm.  The window transparency,
peaking at $\sim 380\,$nm~\cite{EMI}, is approximately matched to the
spectral excitation of a pure CsI fast scintillation light component
with a peak emission at $\sim 310\,$nm at room
temperature~\cite{Woo90}.

Calibrated energy spectra of the calorimeter detector show that the
individual modules receive between 5 and 10\,MeV energy depositions in
a large sample of events averaged over all physics triggers.  The
accumulated radiation doses for individual CsI counters have been
between 40\,rad and 160\,rad, Table~\ref{tab1}.

\begin{figure}[hbt]
\noindent\hbox to \textwidth{\hfill
   \resizebox{0.8\textwidth}{!}
                {\includegraphics{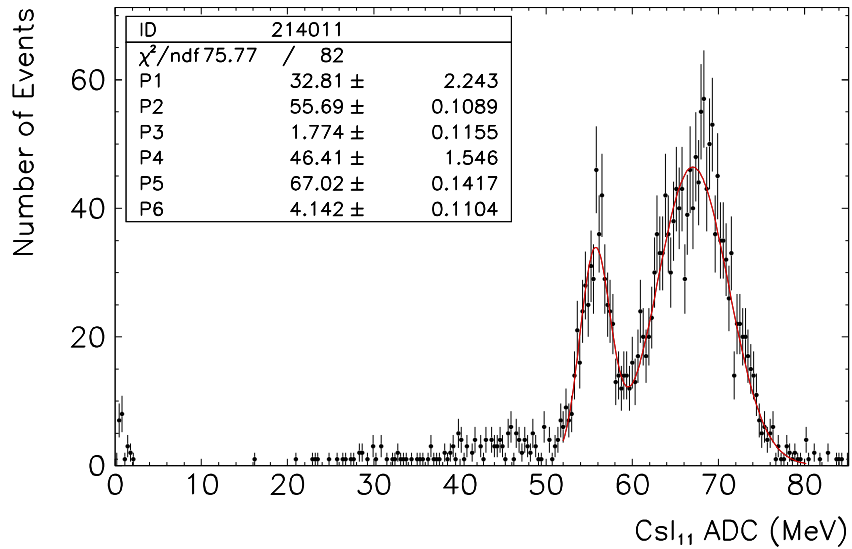}}  \hfill}
\caption{Calibrated energy spectrum of a single CsI detector (CsI
HEX-A module S011, 10 runs).  The higher energy peak at $\sim 68\,$MeV
represents monoenergetic $\pi^+\to e^+\nu$ positrons, the lower part
of the spectrum is populated by $\mu^+\to e^+\nu\bar{\nu}$ Michel
positrons cut by a $\sim 52\,$MeV trigger discriminator threshold.}
\label{fig10}
\end{figure}

A deposited energy (ADC) spectrum, calibrated in MeV and representing
a sum of ADC values for one CsI detector and its nearest neighbors, is
shown as an example in Fig.~\ref{fig10}.  The event trigger is a high
threshold ($\simeq \,52\,$MeV) delayed $\pi^+$ gate trigger that
preferentially selects both $\mu\to e\nu\bar{\nu}$ positrons which
have a continuous energy spectrum with a 52.5\,MeV endpoint, and the
monoenergetic 70\,MeV $\pi\to e\nu$ positrons.  The $\mu$ decay
positron spectrum is clipped on the left very near its endpoint by the
$\simeq \,52\,$MeV trigger discriminator threshold.  The monoenergetic
$\pi$ decay positron line is used for the automatic,
computer-controlled gain matching of the individual CsI detectors.

\begin{figure}[hbt]
\noindent\hbox to \textwidth{\hfill
   \resizebox{0.8\textwidth}{!}
                {\includegraphics{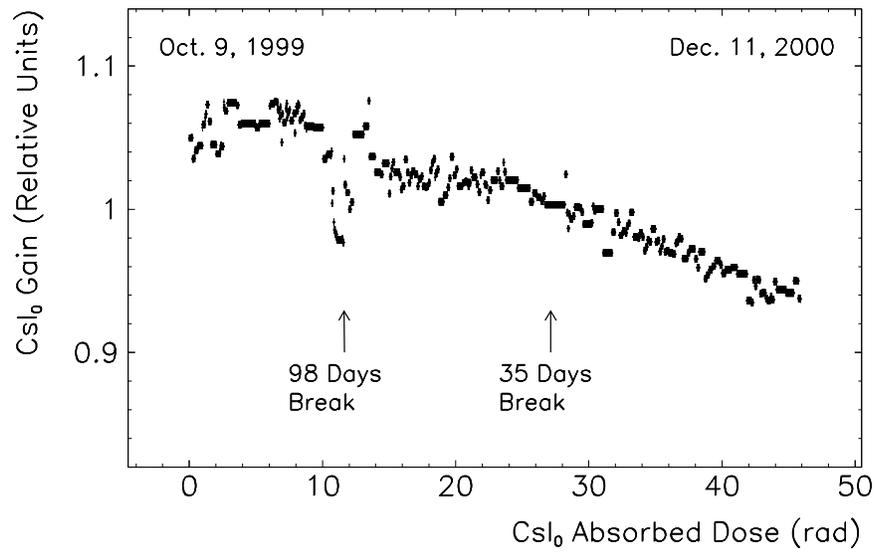}}  \hfill}
\caption{Normalized gain of the CsI detector number 0 (Bicron pentagon
module) as a function of the absorbed radiation dose in rad.}
\label{fig11}
\end{figure}

\begin{figure}[hbt]
\noindent\hbox to \textwidth{\hfill
   \resizebox{0.8\textwidth}{!}
                {\includegraphics{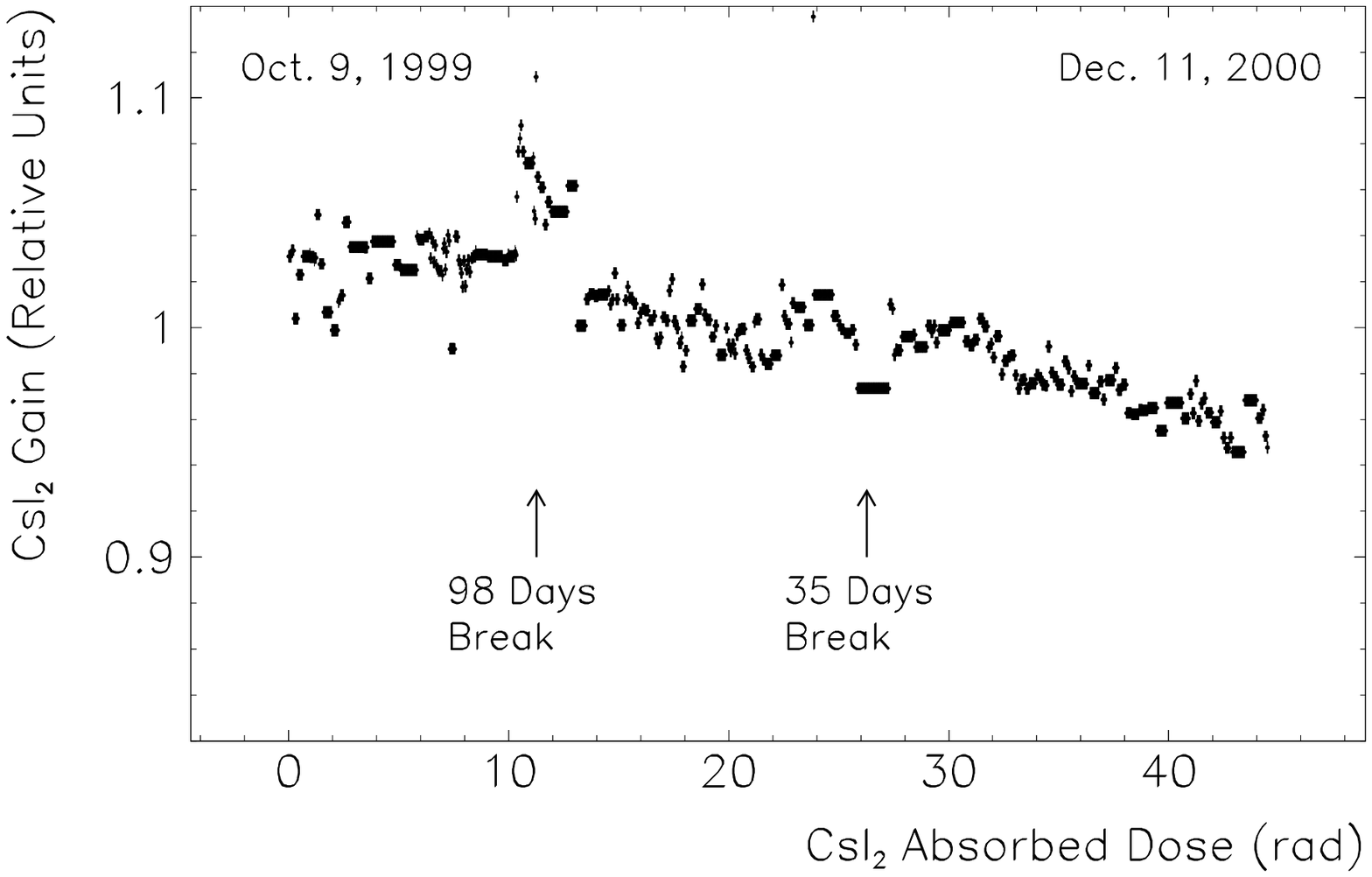}}  \hfill}
\caption{Normalized gain of the CsI detector number 2 (Harkov
pentagon module) as a function of the absorbed radiation dose in rad.}
\label{fig12}
\end{figure}

\begin{figure}[hbt]
\noindent\hbox to \textwidth{\hfill
   \resizebox{0.8\textwidth}{!}
                {\includegraphics{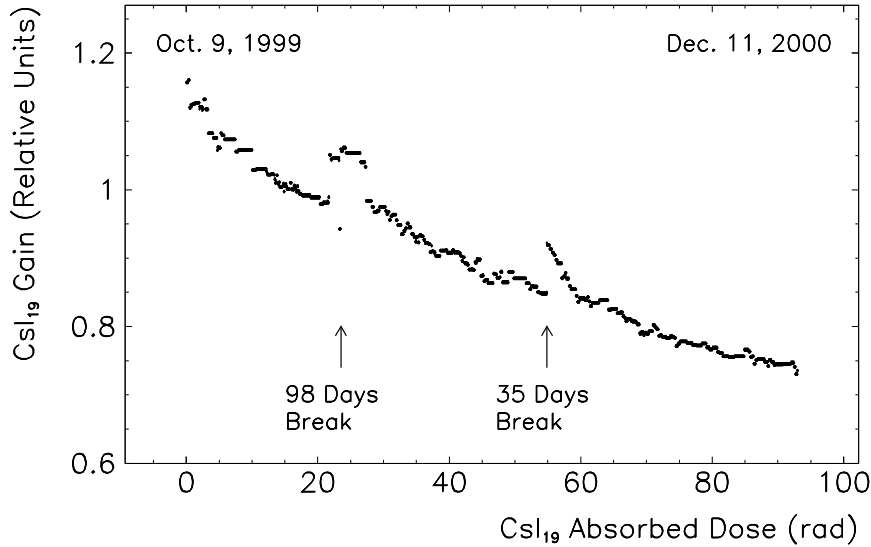}}  \hfill}
\caption{Normalized gain factor of the CsI detector number 19 (Harkov
HEX-A module) as a function of the absorbed radiation dose in rad.}
\label{fig13}
\end{figure}

Because both the software gain factors and the PMT high voltages of
all CsI counters are varied after each run, the relevant real gain
factors are defined by Eq.~(\ref{eq:gain}).  Representative gain
variations for three different CsI detectors are shown in
Figs.~\ref{fig11}--\ref{fig13}.  These measurements show different
gain degradation rates of up to $\ge 40\,$\%/rad.  An increase in gain
following extended beam breaks is a common feature of the CsI light
response, as seen in the three figures.

\clearpage

The overall CsI calorimeter gain and energy resolution are determined
from calibrated experimental energy spectra of all 240 CsI detectors.
The results are summarized in Table~\ref{tab1}.  As it accumulated an
average radiation dose of 119\,rad the PIBETA calorimeter gain
decreased 17\,\% and the overall energy resolution was lowered from
5.5\,\% to 6.0\,\%.  This decrease, however, should not be attributed
entirely to radiation damage, as the crystals were subject to other
sources of performance degradation over time: absorption of humidity
on the crystal surfaces, slow plastic deformation of individual
modules under pressure (CsI is highly malleable, while the detectors
are self-supporting, with the entire weight of over 1\,ton distributed
among the elements).  Thus, the above performance degradation provides
a very loose upper limit on the radiation effects on pure CsI.

\section{Conclusions}

We have analyzed the temporal changes in gains, energy resolutions and
detection efficiencies of the active elements of the PIBETA
detector during 15 month period of production data taking. We find
measurable decreases in energy gains and degradations in energy
resolutions, in particular for the beam detectors that were exposed to
high radiation doses as well as for more radiation-sensitive pure CsI
calorimeter modules.  These changes in the detector responses are
monitored online throughout the production running and documented in
the replay data analysis.  The changes affect the energy calibration
of the PIBETA detector elements and have to be properly taken into
account when defining the software energy cuts in the physics analysis
of rare pion and muon decays.
\bigskip

\noindent {\bf Acknowledgements}

This work has been supported by grants from the US National Science
Foundation.


\bibliographystyle{IEEE}

\newpage

\section*{Tables}

\begin{table}[h]
\begin{centering}

\caption{Summary of long term changes in the gain factors and
fractional energy resolution of selected active elements of the PIBETA
Detector.  `Calo' and `PVeto' give the appropriately averaged 
performance of all CsI and PV elements, respectively.}
\label{tab1} 
\vspace*{2ex}
\begin{tabular}{cccccc}\hline\hline \\[-2ex]
Detector& Radiation  & Starting& Final & Starting & Final    \\
        & dose (krads)& gain    & gain  & res. (\%)& res. (\%)\\[1ex]
\hline                                 \\[-2ex]
BC          & 2000  & 1.00 & 0.66 & 13.1 & 13.9 \\
AD          & 1400  & 1.00 & 0.75 &  7.8 & 8.5  \\
AT$_0$      &  560  & 1.00 & 0.76 &  7.2 & 10.8 \\
AT$_1$      &  630  & 1.00 & 0.89 &  8.4 & 8.6  \\
AT$_5$      &  150  & 1.00 & 0.74 &  7.7 & 8.9  \\
PV$_0$      &   44  & 1.00 & 0.99 & 31.5 & 32.1 \\
PV$_1$      &   40  & 1.00 & 0.99 & 27.6 & 30.4 \\
PVeto       &   41  & 1.00 & 0.95 & 26.2 & 27.9 \\
CsI$_0$     & 0.046 & 1.00 & 0.89 &  5.1 &  5.5 \\
CsI$_2$     & 0.045 & 1.00 & 0.92 &  4.9 &  5.2 \\
CsI$_{11}$  & 0.117 & 1.00 & 1.02 &  6.0 &  6.1 \\
CsI$_{19}$  & 0.093 & 1.00 & 0.65 &  5.0 &  5.3 \\
CsI$_{102}$ & 0.152 & 1.00 & 0.86 &  6.0 &  6.5 \\
CsI$_{165}$ & 0.126 & 1.00 & 0.74 &  5.4 &  5.8 \\
Calo        & 0.119 & 1.00 & 0.83 &  5.5 &  6.0 \\[1ex]
\hline\hline
\end{tabular}

\vspace*{6ex}

\caption{Summary of long term changes in the MWPC and PV hodoscope
charged particle detection efficiencies.}
\label{tab3}

\vspace*{2ex}
\begin{tabular}{cccc}\hline\hline \\[-2ex]
Detector& Radiation  & Starting & Final    \\
        & dose (krads)& eff. (\%)& eff. (\%)\\[1ex]
\hline                                  \\[-2ex]
MWPC$_1$   & 41 & 95.0  &  95.3 \\
MWPC$_2$   & 41 & 98.8  &  98.1 \\
$\Sigma$PV & 41 & 99.1  &  98.9 \\[1ex]
\hline\hline
\end{tabular}

\end{centering}

\end{table}

\end{document}